# Shear-Exfoliated Phosphorene for Rechargeable Nanoscale Battery

Feng Xu[1,2,*], Binghui Ge[3,*], Jing Chen[4,*], Lin Huo[2], Hongyu Ma[5], Chongyang Zhu[1], Weiwei Xia[1], Huihua Min[1], Zhengrui Li[1], Shengli Li[1], Kaihao Yu[1], Feng Wang[2], Yimei Zhu[2], Lijun Wu[2], Yiping Cui[4] & Litao Sun[1]

[1]SEU-FEI Nano-Pico Center, Key Laboratory of MEMS of the Ministry of Education, Southeast University, Nanjing 210096, China. [2]Condensed Matter Physics & Materials Science Department, Brookhaven National Laboratory, Upton, New York 11973, USA. [3]Institute of Physics, Chinese Academy of Sciences, Beijing 100190, China. [4]School of Electronic Science and Engineering, Nanjing 210096, China. [5]Research Center for Internet of Things, China University of Mining and Technology, Xuzhou 221008, China. *These authors contribute equally to this work. Correspondence and requests for materials should be addressed to F.X. (email: fxu@seu.edu.cn) or L.S. (email: slt@seu.edu.cn).

Discovery of atomically thin black phosphorus (called phosphorene) holds promise to be used as an alternative two-dimensional material to graphene and transition metal dichalcogenides especially as an anode material for lithium-ion batteries (LIBs). However, at present bulk black phosphorus (BP) still suffers from rapid capacity fading that results in poor rechargeable performance. Here, for the first time, we use *in situ* transmission electron microscopy (TEM) to construct nanoscale phosphorene LIBs and visualize the capacity fading mechanism in thick multilayer phosphorene by real time capturing delithiation-induced structural decomposition that reduces electrical conductivity and thus causes irreversibility of lithiated $Li_3P$ phase. We further demonstrate that few-layer phosphorene successfully circumvents the structural decomposition and holds superior structural restorability, even subjected to multi-cycle lithiation/delithiation processes and concomitant huge volume expansion. This finding affords new experimental insights into thickness-dependent lithium diffusion kinetics in phosphorene. Additionally, a scalable liquid-phase shear exfoliation route has been developed to produce high-quality ultrathin (monolayer or few-layer) phosphorene, only by a high-speed shear mixer or even a household kitchen blender with the shear rate threshold of $\sim 1.25 \times 10^4\ s^{-1}$, which will pave the way for potential large-scale applications in LIBs once the rechargeable phosphorene nanoscale batteries can be transferred to industrialized enlargement in the future.

## INTRODUCTION

Atomically thin two-dimensional (2D) nanomaterials such as graphene and transition metal dichalcogenides (TMDs) have shown unexpected potential application as an anode material for rechargeable lithium-ion batteries (LIBs) due to their large surface areas and outstanding electrical properties[1–10]. However, these materials inherently possess moderate theoretical capacities, thus there remains a significant demand for other alternative 2D anode materials. Phosphorene, a monolayer or few-layer black phosphorus (BP) atomic layer(s)[11–15], has emerged as a promising candidate to 2D materials due to high theoretical capacity (2596 mAh·g⁻¹)[16,17], high carrier mobility (100–1000 cm²·V⁻¹·s⁻¹)[12,18–20], and remarkable in-plane anisotropic properties[14,21–25]. Although much effort has been devoted to theoretical studies of phosphorene LIBs including lithium diffusion kinetics and structural stability[26–29], no experimental works have been reported so far. At present, only the bulk counterpart of phosphorene, BP, has been experimentally studied as an anode material for LIBs[17,30,31]. But, huge capacity fading in the first discharge/charge cycle resulted in poor rechargeable performance and limited its commercialization[30,31]. The underlying failure mechanism behind the rapid capacity fading still is not understood fully. Therefore, more in-depth works of excavating the failure mechanism experimentally will be greatly appreciated.

Although staying in theoretical studies for phosphorene LIBs[26–29], developing scalable methods in advance to produce high-quality phosphorene still shouldn't be overlooked. Structurally similar to graphite, BP atomic layers are stacked together by weak interlayer forces with significant van der Waals character[32,33], which makes micromechanical exfoliation (Scotch tape delamination) a reliable laboratory-level technique to yield small-scale high-quality phosphorene[34–38]. Large quantities of other 2D nanomaterials such as graphene and TMDs can be produced by chemical[39–43] or electrochemical[44–46] exfoliation, but this approach typically introduces defects or phase transformations that degrade electronic properties, thus not suitable for fabricating high-quality phosphorene. Alternatively, liquid-phase exfoliation by virtue of ultrasonication energy has been proved to be a viable option to prepare high-quality, electronic-grade phosphorene without intermediate chemical reactions[47–51]. However, the scalability of the method is limited by the utilization of sonication as an energy source[52].

In this report, we employed the shear exfoliation technique[52–55] to produce highly crystalline monolayer or few-layer phosphorene nanoflakes by using shear force to break down the interlayer van der Waals forces in appropriate solvents. As-exfoliated half-finished dispersions of phosphorene nanoflakes were subjected to further strict purification treatment. The structure, chemistry, and electrical property of the purified phosphorene nanoflakes are characterized through a comprehensive suite of microscopic and spectroscopic analysis including energy-dispersive X-ray spectroscopy (EDS), high-resolution transmission electron microscopy (TEM), high-angle annular dark field scanning transmission electron microscopy (HAADF-STEM), atomic force microscopy (AFM), scanning electron microscopy (SEM), Raman spectroscopy, and electron energy-loss spectroscopy (EELS). By all of these metrics, the shear-exfoliated 2D phosphorene nanoflakes exhibit competitive properties to mechanically exfoliated counterparts[12,34]. To the best of our knowledge this is the first report of producing phosphorene nanoflakes by shear exfoliation. Importantly, we use *in situ* transmission electron microscopy (TEM) to construct nanoscale phosphorene LIBs and visualize the aforementioned capacity fading mechanism of thick multilayer phosphorene anodes by real time capturing delithiation-induced structural decomposition that reduces electrical conductivity and thus causes irreversibility of lithiated Li₃P phase within the phosphorene matrix. So far there are still no published reports on the lithiation/delithiation behaviors of phosphorene nanoflakes by *in situ* TEM. We further demonstrate that few-layer phosphorene successfully circumvents the structural decomposition and holds superior structural restorability, even subjected to multi-cycle lithiation/delithiation processes and concomitant huge volume expansion. This finding affords new experimental insights into thickness-dependent lithium diffusion kinetics in phosphorene. Overall, our scalable production method of high-quality phosphorene and fundamental understanding of phosphorene lithiation/delithiation behaviors have make the pave for potential large-scale applications of rechargeable phosphorene LIBs.

## RESULTS AND DISCUDDION

Bulk BP crystals are exfoliated in organic solvents that are detailed in the Methods section. The whole exfoliation procedure is carried out in a glovebox with argon gas using a mixing head where a 4-blade rotor sits within a fixed screen known as the stator, as illustrated in Figure 1a. Typical features of the mixing head are narrow gap ($d_{gap}$ = 0.2 mm) between the rotor and stator, and high rotor speed ($N$) creating a high shear rate ($\gamma$) within the gap. Figure 1b show a schematic of BP shear-exfoliation process in the laminar flow regimes that typically occur within the gaps. In a representative experiment, a handful of ground BP powder (0.05 mg/mL) was immersed into *N*-methyl-pyrrolidone (NMP) in a conical tube (Figure 1a). After the shear exfoliation with predetermined time and rotation speed, turbid dispersion (Figure 1b) was obtained that was further purified by centrifugation to remove larger unexfoliated BP crystals. The resulting stable phosphorene dispersions display a brown to pale yellow color as a function of centrifugal rotation speed, as seen in Figure 1c. Additionally, turbulence-assisted shear rate from a household kitchen blender also can be utilized to exfoliate the bulk BP (This section will be expatiated hereinafter).

Following the shear exfoliation and centrifugation in NMP, the phosphorene dispersions were further filtrated by polytetrafluoroethylene (PTFE) membrane and repeatedly rinsed by isopropyl alcohol (IPA) to remove NMP residual. The stacked phosphorene nanoflakes on PTFE membrane (Figure S1a) were renewedly dispersed in IPA for subsequent microscopic and spectroscopic characterizations. Figure 2a shows a typical HAADF-STEM image of an ultrathin phosphorene nanoflake on a lacy carbon support, revealing its uniformity in thickness. The nanoflake is corresponding to the product centrifugated by 6000 rpm rotation speed (see Figure 1c). Its chemical purity is inspected by the EDS analysis. The EDS result (Figure S2) shows that the nanoflake only consist of phosphorus. Comparison between the EDS mapping images (Figure 2b,c) and HAADF-STEM image (Figure 2a) also indicates the high purity of the nanoflake after thoroughly rinsing the shear-exfoliated products, no obvious C element from residual organic solvent is found on the surface of the nanoflake. Moreover, the chemical quality of the nanoflake is further checked by EELS (Figure 2d,e). The invisible plasmon peak in low-loss EELS (Figure 2d) proves that the nanoflake is so thin that plasmon peak signal can't be detected[57]. The characteristic core-loss peak at 133 eV corresponding to the P $L_{2,3}$ edge is observed clearly (Figure 2e)[48]. Absence of the signal $P_xO_y$ peak loss confirms the phosphorus elementary substance all over the whole nanoflake.

The schematic showing the atomic structure of BP crystal along side-view and top-view projections is illustrated in Figure 2f. The monolayer phosphorene consists of P atoms stacked in puckered subplanes, and each P atom is bonded with two adjacent atoms lying in the same plane and with one P atom from another different plane[26]. Here, the monolayer phosphorene is clearly resolved for the phosphorene nanoflake of Figure 2a by high-resolution TEM (HRTEM) imaging and corresponding SAED pattern along [010] projection, as reflected in Figure 2g,h. No noticeable structure defects are observed in the HRTEM image. A filtered section of the HRTEM image (upper right inset in Figure 2g) shows atomic-level structure characteristic of monolayer phosphorene, which obviously is different from the ABA stacking order of multilayer phosphorene (Figure S3). Moreover, we also used XHREM software to perform TEM simulations (viewing from the (001) direction) on monolayer and multilayer phosphorene structures (Figure S3). The simulated monolayer structure (lower right inset in Figure 2g) perfectly matches the imaged HRTEM structure, further corroborating the successful exfoliation of monolayer phosphorene. These TEM analyses provide strong evidence that high-quality monolayer phosphorene with highly crystalline nature can be produced by shear exfoliation. This conclusion also is corroborated by Raman spectroscopy, as given in Figure S4. Four modes are observed at 363 cm$^{-1}$, 438 cm$^{-1}$, 465 cm$^{-1}$, and 523 cm$^{-1}$ assigned to the $A_g^1$, $B_{2g}$, and $A_g^2$ photon modes for phosphorene nanoflakes and the TO phonon mode for the Si substrate, respectively. The Raman spectrum is well in accord with the previous results of phosphorene nanoflakes produced by mechanical exfoliation.

For further AFM analysis of isolated phosphorene nanoflakes, samples were prepared in an Ar glovebox by spin-coating the phosphorene dispersions in IPA onto 200 nm $SiO_2$-coated silicon substrates and then were stored in Ar gas-filled sealing bags until AFM characterization. In Figure 3a, an individual phosphorene nanoflake is imaged by AFM equipped with an environmental cell with dry, ultrahigh purity $N_2$ gas. A step height of ~0.85 nm measured at the nanoflake edge confirms the presence of monolayer phosphorene. No obvious bubbles, droplets, or other signs of phosphorene degradation are found. Even though the step height is slightly larger than the theoretical value of 0.52 nm for monolayer phosphorene, we generally consider that the AFM-measured thickness value of a monolayer 2D nanoflake is larger than the theoretical value due to the roughness height of $SiO_2$/Si substrate; this phenomenon is frequently observed in graphene and TMDs' cases. In addition, phosphorene nanoflakes with step height of ~0.9–3.5 nm also are observed (Figure S5), indicating coexistence of monolayer and few-layer phosphorene nanoflakes produced by the shear exfoliation. The lateral dimension and size of nanoflakes after filtration by PTFE membrane are found to be up to tens of micrometers, as shown in SEM images of Figure 3b–d and Figure S1b–d.

The mechanism of shear-exfoliating graphite has been well expounded by Coleman and colleagues. Just as its name implies, the shear rate ($\gamma$) is responsible for exfoliation. The equation of shear rate can be written as: $\gamma \approx \pi \cdot N \cdot (2R_r)/d_{gap}$, where $N$, $R_r$, $d_{gap}$ are rotor speed, rotor radius, and rotor-stator gap, respectively. Obviously, the shear rate increases proportionally with the rotor speed provided that the mixing head of rotor-stator is fixed ($R_r = 16$ mm, $d_{gap} = 0.2$ mm). Interestingly, we found from TEM characterizations that large quantities of high-quality phosphorene nanoflakes were produced only above a minimum rotor speed of ~1500 rpm. This implies that a minimum shear rate of ~$1.25 \times 10^4$ s$^{-1}$ is a general requirement in our experiments. According to calculation equation of Reynolds-number, $Re = N \cdot (2R_r)^2 \cdot \rho / \eta$, where $\rho$ and $\eta$ is the volume and viscosity of the solvent NMP, respectively, the minimum shear rate can be obtained within the laminar flow regime ($Re < 10^4$), showing turbulence flow ($Re > 10^4$) not to be compulsory for shear exfoliation. The exfoliation process in the laminar flow regime is schematically shown in Figure 1b by modeling exfoliation as shear stress ($F$)-induced interlayer sliding in NMP. According to the previous reports, the same exfoliation mechanism should take place not only in laminar regime but also in turbulent regime. This means that any high-speed mixing head that generates the shear rate higher than ~$1.25 \times 10^4$ s$^{-1}$ is suitable for BP exfoliation. Here, we further demonstrate that phosphorene nanoflakes also can be successfully produced using a Jiuyang kitchen blender (Figure S6a,b and Movie S1). Its rotating blades can generate rotation speed of ~15500–22000 rpm that would result in fully developed turbulence ($Re > 10^4$). Assuming a rated power input of 250 W is completely dissipated via turbulence, the shear rate can be calculated from $\gamma \approx (P/(V \cdot \eta))^{0.5}$, where $V$ and $\eta$ is the volume and viscosity of the solvent NMP, respectively. When 200 mL of NMP was used, the $\gamma$ is $2.73 \times 10^4$ s$^{-1}$ which is larger than the shear rate threshold of ~$1.25 \times 10^4$ s$^{-1}$ for BP delamination. As-exfoliated phosphorene nanoflakes were analyzed by Raman spectroscopy (Figure S6c), TEM imaging (Figure S6d–f), and EELS (Figure S7), and no perceptible differences in flake morphology, purity, and degree of crystallinity are found compare with the aforementioned product prepared by the rotor-stator system. We also note that these nanoflakes are virtually indistinguishable from the sonication-exfoliated counterparts both in terms of size and quality[47–51], indicating that well-exfoliated, high-purity, highly crystalline phosphorene nanoflakes can be produced using a broad range of shear exfoliation conditions.

Bulk BP has been demonstrated as a high-performance anode material in lithium-ion battery (LIB). However, the lithiation-induced tensile stress and layered structure cracking in bulk BP limit its commercialization. The structural changes and stresses may be mitigated by using 1D nanowires or 2D nanoflakes to optimize the structural geometry of anode materials. At present, much effort mainly focuses on theoretical studies of monolayer or few-layer phosphorene as LIB anode material. So far, no experimental works have been conducted to investigate electrochemical lithiation behavior of phosphorene nanoflakes at the nanoscale. Here, our high-quality shear-exfoliated nanoflakes provide the possibility of *in situ* observing the lithiation-induced morphological and structural evolution via TEM, and of unraveling lithium diffusion and reaction kinetics mechanism at the nanoscale. Figure 4a schematically illustrates the *in situ* lithiation experiment of an individual phosphorene nanoflake via a TEM-STM sample holder (Nanofactory Instruments) with a biasing system (Figure 4b). The nanoflakes were glued to the half copper grid with conductive epoxy and then were touched to a sharp tungsten tip that was associated to a piezo-driven biasing-probe, forming an electrical circuit by applying a bias voltage. A lithium metal attached to a tungsten tip was served as the counter electrode and the naturally grown Li$_2$O layer on the surface of Li metal was used as a solid-state electrolyte to allow the diffusion of Li$^+$ ions.

Figure 4c shows the TEM image of the nano-LIB constructed with a slight thick phosphorene nanoflake inside the TEM. The pristine nanoflake is corresponding to the product centrifugated by 1000 rpm rotation speed (see Figure 1c). Its structure can be identified as orthorhombic-phase BP (JCPDS no. 73-1358, space group *Cmca* (64)), as indicated in its single-crystalline SAED pattern (Figure 4d). A typical lithiation experiment was initiated by applying a constant potential of −1 V to the phosphorene nanoflake (as the nano-LIB's anode) with respect to the Li counter electrode to drive the electrons and Li$^+$ ions flow across the circuit after the two electrodes were contacted. Figure 4e–g show time-resolved TEM images from video frames (Movie S2) during the electrochemical lithiation process. It's found that the nanoflake longitudinally elongated from 1.7 μm to 2.3 μm, with an expansion rate of 36%. The nearly same expansion rate of 35% from 1 μm to 1.35 μm also was observed for the lateral direction, which implies the same diffusion rate of Li$^+$ ions at different directions. The isotropic lithiation behavior isn't consistent with the theoretical studies that reveal the anisotropic lithium diffusion on the surface of monolayer phosphorene. Especially, density functional theory (DFT) calculations by Zhang, *et al.* concluded that the shallow energy barrier (0.08 eV) of Li diffusion on monolayer phosphorene along zigzag direction ([100] direction) leaded to an ultrahigh diffusivity, which was estimated to be 10$^2$ (10$^4$) times faster than that on MoS$_2$ (graphene). On the contrary, the large energy barrier (0.68 eV) along the armchair direction ([001] direction) resulted in a nearly forbidden diffusion. It is worthy to note that the DFT calculations further point out that the huge difference in diffusion energy barrier along the zigzag and armchair directions could remarkably be reduced for multilayer phosphorene nanoflakes, which can explain the observed isotropic lithiation of the slightly thick phosphorene nanoflake. Monitoring microstructure evolution of the nanoflake affords the opportunity to better understand the lithiation process and phase transition. After lithiation, the SAED exhibits the poly-crystalline diffraction rings with discontinuous circlewise spots (Figure 4h) that can be indexed as hexagonal-phase Li$_3$P (JCPDS no. 74-1160, space group *P63/mmc* (194)). The phase transition from P to Li$_3$P and concomitant volume expansion also have been found in bulk BP LIB anodes, which can provide a relatively large theoretical capacity.

The electrochemical delithiation behavior also was *in situ* inspected by applying a reversed constant potential of 1 V, as shown in Figure 4i–k and Movie S3. Interestingly, the anticipated dimensional shrinkage associated with the deintercalation of Li$^+$ ions didn't take place for the nanoflake; instead, the nanoflake suddenly decomposed. The decomposition began from one end contacting lithium counter electrode (Figure 4j) and then rapidly propagated to the other end (Figure 4k), forming discrete regions in the nanoflake matrix. A mass of residual lithium in the formation of Li$_3$P phase still existed in the delithiated nanoflake, as reflected in the SAED pattern (Figure 4l) and EELS data (Figure S8). The discrete regions and irreversible Li$_3$P phase in the delithiated nanoflake possibly would result in loss of electrical contact and thus rapid capacity fading. For instance, the huge capacity fading of ~43% in the first lithiation/delithiation cycle has been found for the bulk BP LIB anodes, which impeded successful application of bulk BP anodes. In this regard, the compromised reaction of P↔LiP rather than P↔Li$_3$P by controlling the voltage range was used to provide a stable but low specific capacity (~600 mAh·g$^{-1}$). Thus, stable rechargeable cycle has not yet been achieved when accessing the full capacity (at least ~2000 mAh·g$^{-1}$) of bulk BP anodes. Here, for the first time, we revealed the capacity fading mechanism associated with the delithiation-induced structural decomposition at the nanoscale. This finding affords us new insight into redesigning the bulk BP anodes. The anisotropic lithiation behavior and structural decomposition phenomenon were repeatedly corroborated by additional *in situ* lithiation/delithiation experiments, as shown in Figure S9 and Movie S4.

Fortunately, we have found that few-layer phosphorene nanoflakes can keep structural integrity even subjected to multi-cycle lithiation/delithiation processes (Figure 5 and Movie S5–7). Here we use the centrifugation at 6000 rpm rotation speed to screen out few-layer phosphorene nanoflakes (see Figure 1c). After applying a constant potential of −1 V to an individual few-layer phosphorene nanoflake with respect to the Li counter electrode, the lithiation resulted in a lateral inflation from 1.9 μm to 2.4 μm with an expansion rate of 26% (Figure 5a–c and Movie S5). However, the longitudinal expansion didn't take place, which implies different diffusion rates of $Li^+$ ions at different directions (such as zigzag direction and armchair direction). The anisotropic lithiation behavior is well consistent with the DFT theoretical studies concerning the anisotropic lithium diffusion on monolayer phosphorene. According to DFT calculations, the large energy barrier (0.68 eV) of Li diffusion on monolayer phosphorene along the armchair direction resulted in a nearly forbidden diffusion compared with an ultrahigh Li diffusivity along zigzag direction with the shallow energy barrier (0.08 eV). In a way, the theoretical result also can account for the anisotropic lithiation behavior in our *in situ* experiments, proving that the exfoliated few-layer phosphorene is comparable to monolayer phosphorene.

Excitingly, the nanoflake could completely restore its original morphology and size (Figure 5d–f and Movie S5) after the delithiation, and we didn't observe any visible structural decomposition that had occurred in the thick phosphorene nanoflakes. Significantly, the nearly same expansion rate and restorability in morphology were kept after the second lithiation/delithiation cycle, as shown in Figure 5g–l and Movie S6. Even after dozens of lithiation/delithiation cycles, the few-layer nanoflake still can restore its original morphology and size provided that the applied potential is fixed. When a higher potential of −3 V was used to lithiate the nanoflake, a larger lateral expansion rate of 45% was obtained (Figure 5m–o and Movie S7). The longitudinal expansion wasn't found. The large expansion yet didn't result in the structural decomposition of the nanoflake during the delithiation by applying an equivalent reversed potential (Figure 5p–r and Movie S7). The structural stability also was verified by dozens of lithiation/delithiation cycles. This finding provides powerful evidence that the few-layer phosphorene nanoflakes exhibit better electrochemical cycleability compared with the bulk counterpart BP as the LIB anodes, and possess the great commercialization prospect in high-performance LIBs.

**Acknowledgements**

This work was supported by the National Basic Research Program of China (973 Program, Grant No. 2015CB352106), the National Natural Science Foundation of China (NSFC, Grant Nos. 51372039, 61574034, 51202028), the Jiangsu Province Science and Technology Support Program (Grant No. BK20141118), the Fundamental Research Funds for the Central Universities (Grant Nos. 2242013R30004), China Postdoctoral Science Foundation Funded Project (Grant No. 2014M550259, 2015T80480). The work at Brookhaven National Lab is supported by U.S. DOE-BES under Contract number DE-AC02-98CH10886.

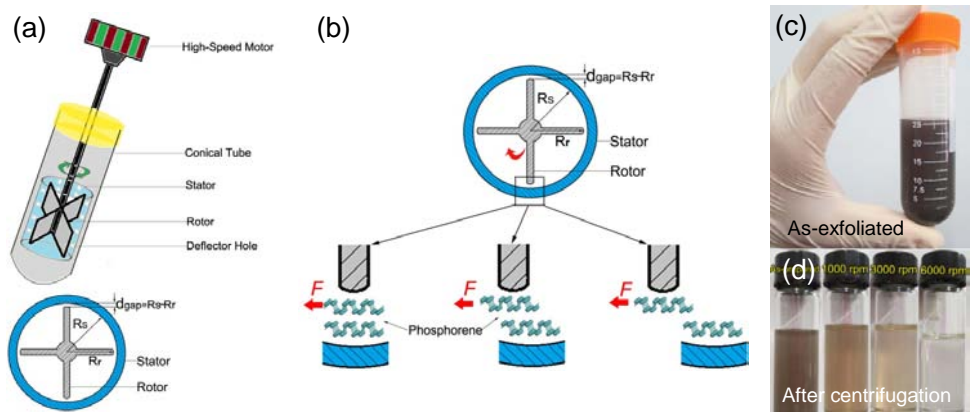

Figure 1 (a) Schematic of shear exfoliation setup for exfoliating bulk BP. (b) Schematic representing the shear-exfoliation process in the laminar flow regime. (c) Photograph of as-exfoliated phosphorene nanoflakes dispersed in NMP before centrifugation. (d) Photograph of phosphorene nanoflakes dispersed in NMP after 1000 rpm, 3000 rpm and 6000 rpm centrifugation, respectively.

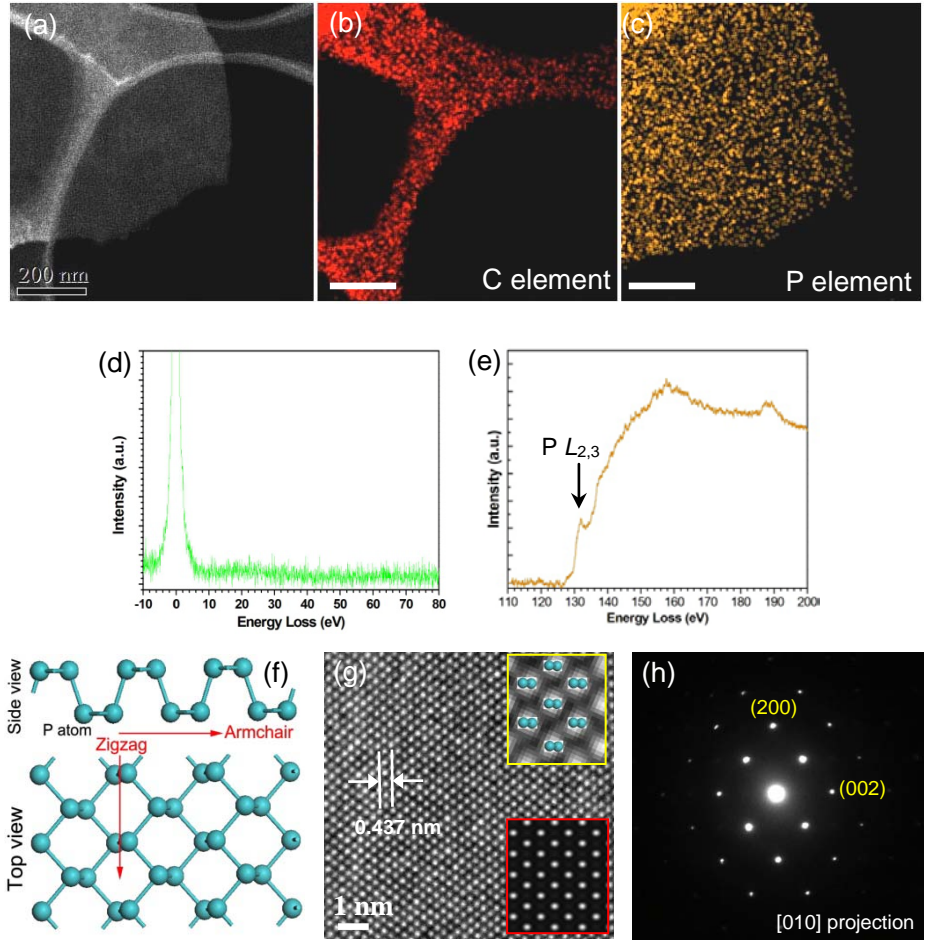

Figure 2 HAADF-STEM, EDS mapping, SAED, EELS characterization of phosphorene nanoflakes prepared by shear exfoliation and centrifugation at a rotation of 6000 rpm. (a) A typical STEM-HAADF image of an ultrathin phosphorene nanoflake on a lacy carbon support (scale bar is 200 nm), revealing its uniformity in thickness. (b, c) EDS mapping of C and P elements corresponding to (a). Comparison between STEM-HAADF image and element mapping images indicates the high-purity of the phosphorene nanoflake after thoroughly filtrating and rinsing the shear-exfoliated products, no obvious C element from residual organic solvent is found on the surface of the nanoflake. (d) and (e) are low-loss and core-loss EELS of the phosphorene nanoflake. The characteristic core-loss peak at 133 eV corresponding to the P $L_{2,3}$ edge is observed clearly. Absence of the signature $P_xO_y$ peak loss confirms the pure phosphorus structure all over the nanoflake. The invisible plasmon peak in low-loss EELS (d) further proves that the nanoflake is extremely thin. (f) Schematic showing the atomic structure of phosphorene along side view and top view. (g) High-resolution TEM (HRTEM) image and (h) corresponding SAED pattern of the monolayer phosphorene corresponding to (a). The projection direction is along [010]. This HRTEM analysis provides strong evidence that high-quality monolayer phosphorene with highly crystalline nature can be produced by shear exfoliation. A filtered section of the HRTEM image (upper right inset in (g)) and TEM image simulation of monolayer phosphorene (lower right inset in (g)) are added for comparison.

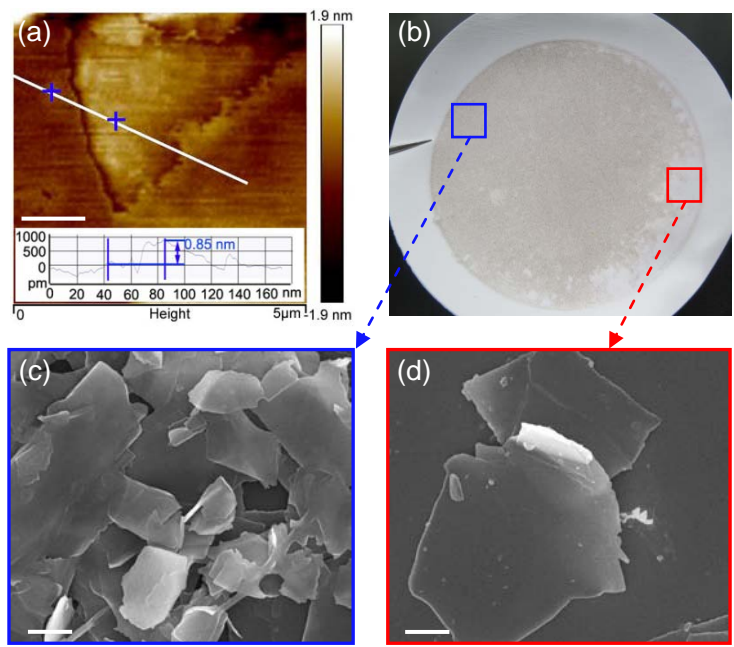

**Figure 3** AFM and SEM characterization of shear-exfoliated phosphorene nanoflakes. (a) AFM height measurement of a monolayer phosphorene nanoflake on a 200 nm $SiO_2$-coated Si substrate. The height profile (inset) corresponds to the drawn white line. The stage height of around 0.85 nm proves the existence of a monolayer phosphorene nanoflake. Scale bar is 200 nm. (b) Photograph and (c, d) SEM images of of as-filtrated film of phosphorene nanofalkes on polytetrafluoroethylene (PTFE) membrane, showing different size and dimensionality (scale bar is 2 μm).

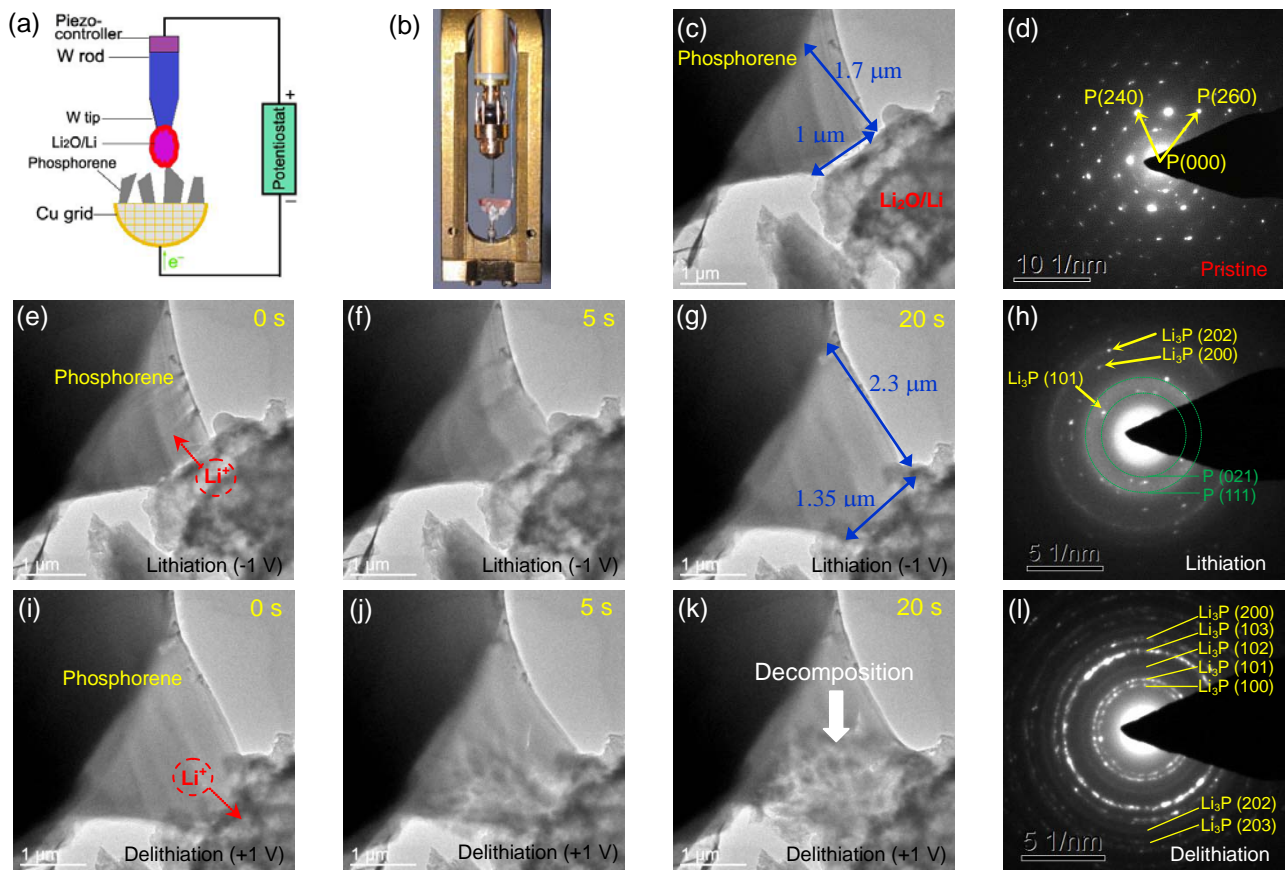

Figure 4 *In situ* TEM electrochemical lithiation/delithiation experiments of a slightly thick phosphorene nanoflake prepared by shear-exfoliation in NMP and centrifugation at a rotation speed of 1000 rpm. (a) Schematic and (b) Photograph of *In situ* lithiation experiment setup of TEM-STM sample holder. (c) TEM image of the nano-LIB constructed with an individual phosphorene nanoflake inside the TEM. (d) SAED pattern of the pristine phosphorene nanoflake, showing BP single-crystalline orthorhombic structure before lithiation. (e–g) SAED pattern of the lithiated phosphorene nanoflake, revealing the formation of $Li_3P$ after lithiation. Time-resolved TEM images from video frames show morphological evolutions of the phosphorene nanoflakes during the electrochemical lithiation process, and a lateral expansion of around 30% was observed for the lithiated phosphorene nanoflakes. (g) (i–k) Time-resolved TEM images from video frames show morphological evolutions of the phosphorene nanoflake during the electrochemical delithiation process. (l) SAED pattern of the delithiated phosphorene nanoflake, revealing $Li_3P$ residual after delithiation.

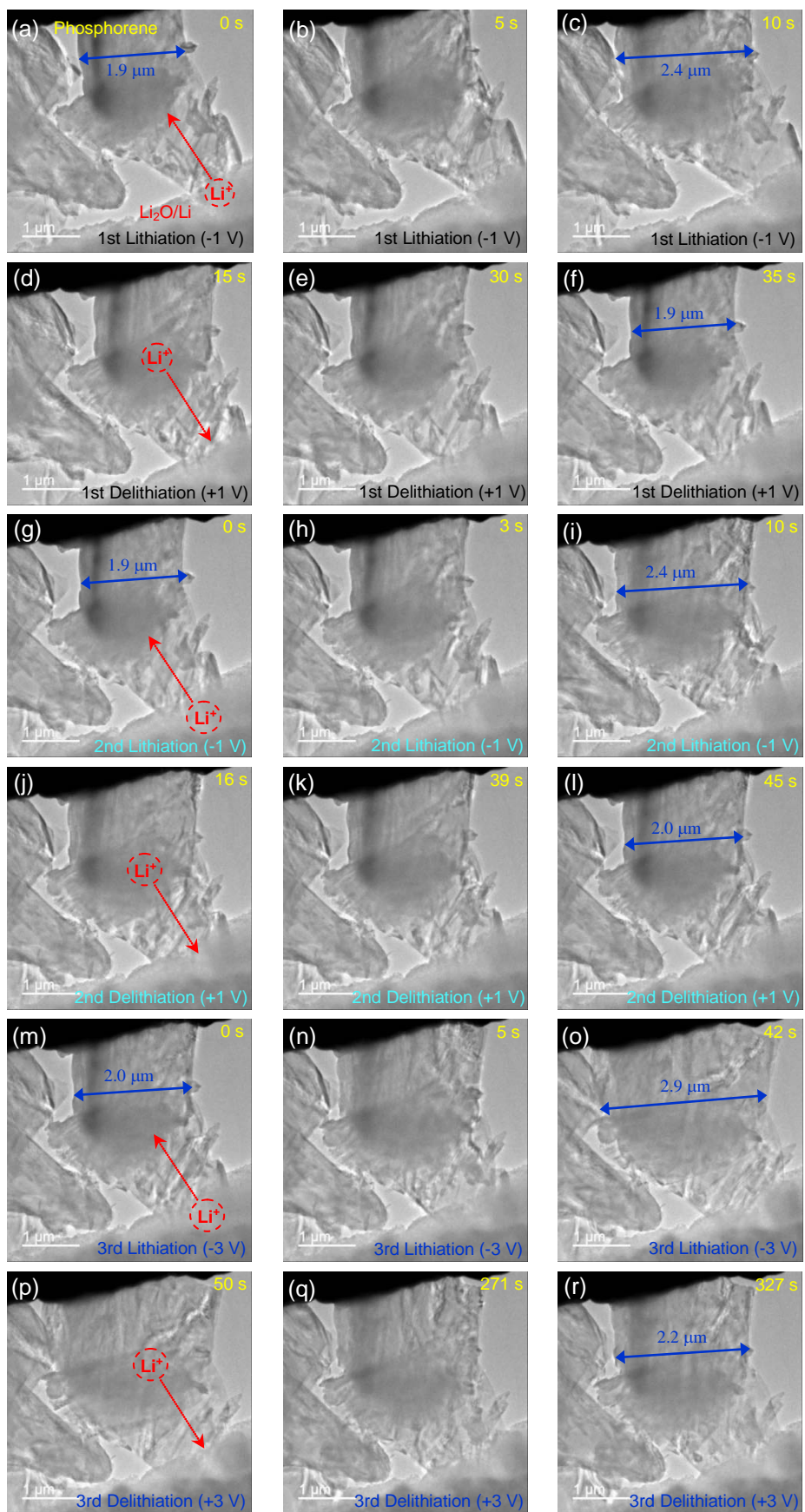

Figure 5 Multi-cycle *in situ* TEM electrochemical lithiation/delithiation experiments of an individual few-layer phosphorene nanoflake prepared by shear-exfoliation in NMP and centrifugation at a rotation speed of 6000 rpm. No decomposition phenomena took place during the delithiation, proving the superior structural stability compared with the thick phosphorene nanoflakes. (a–f) Time-resolved TEM images of the first lithiation/delithiation cycle, showing superior morphological restorability and structural stability. (g–l) Time-resolved TEM images of the second lithiation/delithiation cycle, maintaining the same morphological restorability and structural stability as the first lithiation/delithiation cycle. In both the lithiation/delithiation two cycles, the same potential of −1 V/1 V was applied. (m–r) Time-resolved TEM images of the third lithiation/delithiation cycle with a larger applied potential of −3 V/3 V. Although a larger lateral expansion rate took place, the few-layer nanoflake still kept the superior morphological restorability and structural stability after the delithiation.